# Quantum Phase Transition in the BCS-to-BEC Evolution of p-wave Fermi Gases

S. S. Botelho and C. A. R. Sá de Melo

*School of Physics, Georgia Institute of Technology, Atlanta, GA 30328, USA*

**Abstract.** We discuss the possibility of a quantum phase transition in ultra-cold spin-polarized Fermi gases which exhibit a $p$-wave Feshbach resonance. We show that when fermionic atoms form a condensate that can be externally tuned between the BCS and BEC limits, the zero temperature compressibility and the spin susceptibility of the fermionic gas are non-analytic functions of the two-body bound state energy. This non-analyticity is due to a massive rearrangement of the momentum distribution in the ground state of the system. Furthermore, we show that the low temperature superfluid density is also non-analytic, and exhibits a dramatic change in behavior when the critical value of the bound state energy is crossed.

## 1. INTRODUCTION

Recent experiments in cold fermionic gases have shown that $s$-wave magnetic field induced Feshbach resonances can be used to study the BCS-to-BEC evolution [1, 2, 3, 4, 5, 6] from large Cooper pairs on the higher magnetic field side of the resonance (BCS regime) to small diatomic molecules on the lower magnetic field side of the resonance (BEC regime). These studies led to the first experimental realization of the theoretically proposed BCS-to-BEC crossover in three dimensional continuum $s$-wave superfluids [7, 8]. Three early theoretical works that considered the possibility of $s$-wave superfluidity in the context of (what is known today as) the BCS-to-BEC crossover should be highlighted. The first is by Eagles [9], where the possibility of pairing without condensation is described in a continuum model in the context of superconductors with low carrier concentration [10]. The second is Leggett's seminal work [11], in which the $T = 0$ $s$-wave and $p$-wave BCS-to-BEC evolution are discussed as a crossover phenomenon in the context of a variational ground state wavefunction. And the third is the work of Nozieres-Schmitt-Rink [12], where the $s$-wave BCS-to-BEC crossover in a lattice is described. Furthermore, much of the theoretical [13, 14, 15, 16] and experimental [1, 2, 3, 4, 5, 6] efforts that followed described only the BCS-to-BEC crossover in $s$-wave systems.

In this manuscript, we present a functional integral analysis of the BCS-to-BEC evolution in $p$-wave fully spin-polarized Fermi gases, where $p$-wave Feshbach resonances have already been observed [17, 18]. We show that a quantum phase transition takes place when the chemical potential crosses a critical value, instead of the usual smooth BCS-to-BEC crossover that occurs in $s$-wave superfluids [19]. The atomic compressibility and the spin susceptibility of the Fermi gas are computed and are shown to be non-analytic in the $p$-wave case, as a consequence of a major rearrangement in the momentum distribution as the critical point is approached. This non-analytic behavior suggests the occurrence of a quantum phase transition, which is further confirmed by a discontinuous change in the temperature dependence of the superfluid density of the gas at the transition point, which goes from power-law on the BCS side of the resonance to exponential on the BEC side of the resonance.

We study the case of quasi-two-dimensional systems, which can be prepared experimentally through the formation of a one-dimensional optical lattice, where atom transfer between lattice sites is suppressed by a large trapping potential. The form of the trapping potential can be chosen to be

$$V_{\text{trap}} = -V_0 \exp\left[-2\left(\frac{x^2}{w_x^2} + \frac{y^2}{w_y^2}\right)\right] \cos^2(k_z z), \qquad (1)$$

where $2\pi/k_z$ is the wavelength of the light used in the laser beam and $w_x \neq w_y$, so that the trap is asymmetric in the $x$ and $y$ directions. We also assume that $w_x, w_y \gg \lambda_F$, where $\lambda_F = 2\pi/k_F$ is proportional to the interparticle spacing of a Fermi gas with Fermi wavevector $\mathbf{k}_F$. Finally, we require that $\varepsilon_F \ll V_0 - E_0$, where $\varepsilon_F = \hbar^2 k_F^2/2m$ is the Fermi energy in two dimensions and $E_0$ is the ground state of the Gaussian potential (with respect to its bottom), such that the tunneling between two minima of the trapping potential is essentially suppressed, and the problem can be considered quasi-two-dimensional.

The rest of the manuscript is organized as follows. In Section 2, we discuss the Hamiltonian and the interparticle potential used in our model, and present the order parameter symmetries analyzed in this work. In Section 3, our functional integral calculation is carried out, and the order parameter and number equations are derived at the saddle point level of approximation. Our numerical results for the chemical potential and order parameter amplitude as functions of the binding energy are also shown in this section. The analysis of Gaussian fluctuations about the saddle point solution is performed in Section 4. Then, in Sections 5, 6 and 7, our results for the momentum distribution, atomic compressibility and spin susceptibility are presented. The analysis of the superfluid density is performed in Section 8, while the possibility of a Berezinskii-Kosterlitz-Thouless phase transition in the system under consideration is discussed in Section 9. Finally, our concluding remarks are summarized in Section 10.

## 2. HAMILTONIAN AND INTERACTION POTENTIAL

We study a uniform quasi-two-dimensional continuum model of *spin-polarized* (all atoms in the same hyperfine state) fermionic atoms of mass $m$ and density $n = k_F^2/4\pi$. In the presence of an external magnetic field $\mathbf{h}$, the system is described by the Hamiltonian ($\hbar = k_B = 1$)

$$\mathcal{H} = \sum_{\mathbf{k}} \xi_{\mathbf{k}} \psi^{\dagger}_{\mathbf{k}\uparrow} \psi_{\mathbf{k}\uparrow} + \frac{1}{2} \sum_{\mathbf{k},\mathbf{k}',\mathbf{q}} V_{\mathbf{k}\mathbf{k}'} b^{\dagger}_{\mathbf{k}\mathbf{q}} b_{\mathbf{k}'\mathbf{q}}, \qquad (2)$$

where $b_{\mathbf{kq}} = \psi_{-\mathbf{k}+\mathbf{q}/2\uparrow}\psi_{\mathbf{k}+\mathbf{q}/2\uparrow}$ and $\xi_{\mathbf{k}} = \varepsilon_{\mathbf{k}} - \tilde{\mu}$, with $\varepsilon_{\mathbf{k}} = k^2/2m$, and $\tilde{\mu} = \mu + g_{\tilde{z}\tilde{z}}\mu_B h_{\tilde{z}}$. The direction of the magnetic field $\mathbf{h}$, which was chosen to define the spin quantization axis $\tilde{z}$, need not to coincide with the spatial direction $z$ of the laser beam.

In order to obtain an approximate expression for the atomic interaction potential, we start by using the Fourier expansion of a plane-wave in two dimensions,

$$e^{i\mathbf{k}\cdot\mathbf{r}} = \sum_{\ell=-\infty}^{\infty} i^{\ell} J_{\ell}(kr) e^{i\ell\phi}, \tag{3}$$

where $\phi = \arccos(\mathbf{k}\cdot\mathbf{r})$ and $J_{\ell}(kr)$ is a Bessel function of order $\ell$, into the matrix element of the interaction potential in $k$-space. This leads to

$$V_{\mathbf{kk'}} = \langle \mathbf{k}|V|\mathbf{k'} \rangle = \sum_{\ell=-\infty}^{\infty} e^{i\ell\theta_{\mathbf{kk'}}} V_{kk'}^{(\ell)}, \tag{4}$$

where $\theta_{\mathbf{kk'}} = \arccos(\mathbf{k}\cdot\mathbf{k'})$, and the $k$-dependent coefficients $V_{kk'}^{(\ell)}$ are related to the real space potential $V(r)$ through the Bessel transform,

$$V_{kk'}^{(\ell)} = 2\pi \int_0^{\infty} dr\, r J_{\ell}(kr) J_{\ell}(k'r) V(r). \tag{5}$$

The index $\ell$ labels angular momentum states in two spatial dimensions, with $\ell = \pm 1, \pm 3, \ldots$ corresponding to $p, f, \ldots$ channels, respectively.

In the long-wavelength limit ($k \to 0$), one can show that the $k$-dependence of this potential becomes exactly separable. In fact, for $kr \leq 1$ and $k'r \leq 1$, the asymptotic expression of the Bessel function for small arguments can be used, giving $V_{kk'}^{(\ell)} = C_{\ell} k^{\ell}(k')^{\ell}$, with the coefficient $C_{\ell}$ dependent on the particular choice of the real space potential. In the opposite limit, $kr \gg 1$ and $k'r \gg 1$, the potential is certainly not separable. However, one can show that, in this case, $V_{kk'}^{(\ell)}$ mixes different $k$ and $k'$ and shows an oscillatory behavior which is dependent on the exact form of $V(r)$, with a decaying envelope that is proportional to $k^{-1/2}(k')^{-1/2}$.

Under these circumstances, we choose to study a potential that contains most of the features described above. One possibility is to retain only the $\pm\ell$ terms in Eq.(4), which amounts to isolating only the contribution from the $\ell^{\text{th}}$ angular momentum channel to the scattering process responsible for the interaction between the fermionic atoms. Using the properties discussed above, this results in

$$\begin{aligned} V_{\mathbf{kk'}} &= -\lambda h(k) h(k') \cos(\ell(\varphi - \varphi')) = \\ &= -\lambda h(k) h(k') \left[\cos(\ell\varphi)\cos(\ell\varphi') + \sin(\ell\varphi)\sin(\ell\varphi')\right], \end{aligned} \tag{6}$$

where $\lambda$ is the interaction strength and the function $h(k) = (k/k_1)^{\ell}/(1+k/k_0)^{\ell+1/2}$ controls the range of the interaction, with $R_0 \sim k_0^{-1}$ playing the role of the interaction range, and $k_1$ setting the scale at low momenta. This function indeed reduces to $h(k) \sim k^{\ell}$ for small $k$, and behaves as $h(k) \sim k^{-1/2}$ for large $k$, which guarantees the correct behavior expected for $V_{kk'}^{(\ell)}$ according to the previous analysis.

Another possibility is to keep only the $+\ell$ term in the plane-wave expansion of Eq.(4), which leads to $V_{\mathbf{k}\mathbf{k}'} = -\lambda \Gamma(\mathbf{k})\Gamma^*(\mathbf{k}')$, with $\Gamma(\mathbf{k}) = h(k)e^{i\ell\varphi}$, resulting in a complex order parameter with an angle-independent energy gap. However, as we will show later, one obtains qualitatively equivalent results with this interaction potential in the present quasi-two-dimensional case.

In the strict case of isotropic trapping ($w_x = w_y$), either one of these two options is a plausible possibility. However, the interaction potential becomes exactly separable only in the case where one angular momentum component is selected (either $\ell = +1$ or $\ell = -1$), which will lead to a triplet order parameter without nodes (in the BCS limit), similar to the BW phase of $^3$He in 3D.

On the other hand, if one has either an anisotropic trap ($w_x \neq w_y$) or a weak perturbing asymmetric potential ($V_{\text{add}} = -U_x \cos^2(k_x x) - U_y \cos^2(k_y y)$) on top of a symmetric trap, then the polar symmetry in the xy-plane is broken, and the existence of an order parameter with nodes (in the BCS limit), similar to the ABM phase of $^3$He in 3D, becomes possible. This externally controlled symmetry breaking (via the focus, wavelength, and intensity of the laser beams) may lead to the existence of a spin-polarized superfluid state with a $p_x$ or $p_y$ order parameter. In this spirit, we will choose a phenomenological separable interaction potential that leads to a $p_x$ or $p_y$ order parameter based on the symmetry considerations just mentioned. This can be achieved by going back to Eq.(6) and keeping only one of the terms inside brackets. Assuming, for instance, that the $p_x$-pairing will be favored over the $p_y$-pairing, then only the cosine term is retained, resulting in a fully separable interaction potential in k-space,

$$V_{\mathbf{k}\mathbf{k}'} = -\lambda \Gamma(\mathbf{k})\Gamma(\mathbf{k}'), \tag{7}$$

where $\Gamma(\mathbf{k}) = h(k)\cos(\ell\varphi)$. In particular, for p-wave symmetry ($\ell = 1$), one has

$$\Gamma(\mathbf{k}) = \frac{(k/k_1)}{(1 + k/k_0)^{3/2}} \cos\varphi. \tag{8}$$

Symmetry requirements based on lattice effects and anisotropy have been extensively used in the literature to justify the separation between angular momentum channels, most notably in the study of high-$T_c$ superconductors [20], where the $\ell = 2$ channel in the case of tetragonal symmetry in two-spatial dimensions has only two possible 1D representations: $d_{x^2-y^2}$ and $d_{xy}$. In fact, this idea has already been used in the study of the BCS-to-BEC evolution of d-wave superconductors in 2D [19, 21], in which case a similar interaction potential is considered and symmetry arguments are used to select a $d_{x^2-y^2}$ order parameter over a $d_{xy}$ order parameter.

In the limit of small momenta, this approach is identical to the T-matrix formalism [11], but has the added advantage of making unnecessary to introduce a scattering length as a relevant parameter, which is quite problematic in two-dimensions [22]. The BCS-to-BEC evolution can be safely analyzed provided that the system is dilute enough ($k_F^2 \ll k_0^2$), i.e., the square of the interparticle spacing ($\sim k_F^{-1}$) is much larger than the square of the interaction range ($\sim k_0^{-1}$). Throughout the manuscript, we choose to scale all energies with respect to the Fermi energy $\varepsilon_F = k_F^2/2m$ and all momenta with respect to $k_F$. We will present results for the two cases analyzed, corresponding to the symmetry factors $\Gamma(\mathbf{k}) = h(k)\cos(\varphi)$ (leading to a $p_x$-symmetry order parameter) and $\Gamma(\mathbf{k}) = h(k)e^{\pm i\varphi}$ (leading to a $p_x \pm ip_y$-symmetry order parameter).

## 3. EFFECTIVE ACTION AND SADDLE POINT EQUATIONS

The partition function $Z$ at a temperature $T = \beta^{-1}$ is written as an imaginary-time functional integral with action $S = \int_0^\beta d\tau [\sum_\mathbf{k} \psi_{\mathbf{k}\uparrow}^\dagger(\tau) \partial_\tau \psi_{\mathbf{k}\uparrow}(\tau) + \mathscr{H}]$. Introducing the usual Hubbard-Stratonovich field $\phi_\mathbf{q}(\tau)$, which couples to $\psi^\dagger\psi^\dagger$, and integrating out the fermionic degrees of freedom, we obtain

$$Z = \int \mathscr{D}\phi \mathscr{D}\phi^* \exp(-S_{\text{eff}}[\phi,\phi^*]), \tag{9}$$

with the effective action given by

$$S_{\text{eff}} = \int_0^\beta d\tau \left[ U(\tau) + \sum_{\mathbf{k},\mathbf{k}'} \left( \frac{\xi_\mathbf{k}}{2} \delta_{\mathbf{k},\mathbf{k}'} - \text{Tr}\ln \frac{1}{2}\mathbf{G}_{\mathbf{k},\mathbf{k}'}^{-1}(\tau) \right) \right], \tag{10}$$

where $U(\tau) = \sum_\mathbf{k} |\phi_\mathbf{k}(\tau)|^2/(2\lambda)$ and $\mathbf{G}_{\mathbf{k},\mathbf{k}'}^{-1}(\tau)$ is the (inverse) Nambu matrix,

$$\mathbf{G}_{\mathbf{k},\mathbf{k}'}^{-1}(\tau) = \begin{pmatrix} -(\partial_\tau + \xi_\mathbf{k})\delta_{\mathbf{k},\mathbf{k}'} & \Lambda_{\mathbf{k},\mathbf{k}'}(\tau) \\ \Lambda_{\mathbf{k}',\mathbf{k}}^*(\tau) & -(\partial_\tau - \xi_\mathbf{k})\delta_{\mathbf{k},\mathbf{k}'} \end{pmatrix}, \tag{11}$$

with $\Lambda_{\mathbf{k},\mathbf{k}'}(\tau) = \phi_{\mathbf{k}-\mathbf{k}'}(\tau)\Gamma((\mathbf{k}+\mathbf{k}')/2)$.

*Saddle Point Equation:* After Fourier transforming from imaginary time to Matsubara frequency ($ik_n = i(2n+1)\pi/\beta$) and performing the frequency sum, the saddle point condition $[\delta S_{\text{eff}}/\delta \phi_\mathbf{q}^*(\tau')]_{\Delta_0} = 0$ can be cast in the form of the familiar order parameter equation,

$$\frac{1}{\lambda} = \sum_\mathbf{k} \frac{|\Gamma(\mathbf{k})|^2}{2E_\mathbf{k}} \tanh\left(\frac{\beta E_\mathbf{k}}{2}\right), \tag{12}$$

where $E_\mathbf{k} = \sqrt{\xi_\mathbf{k}^2 + |\Delta_\mathbf{k}|^2}$ is the quasiparticle excitation energy, and $\Delta_\mathbf{k} = \Delta_0 \Gamma(\mathbf{k})$ plays the role of the order parameter function.

At this point, it is interesting to show in some detail how the interaction strength $\lambda$ can be eliminated in favor of the two-body bound state energy $E_b(h_{\tilde{z}})$ in vacuum (and in the presence of a magnetic field). A relation between these two quantities can be otained by solving the Schroedinger equation for two fermions interacting via the pairing potential $V(r)$. After Fourier transforming from center-of-mass coordinates to $k$-space, the Schroedinger equation for the pair wave function $\psi_\mathbf{k}$ becomes

$$2\varepsilon_\mathbf{k} \psi_\mathbf{k} + \sum_{\mathbf{k}'} V_{\mathbf{k}\mathbf{k}'} \psi_{\mathbf{k}'} = \tilde{E}_b \psi_\mathbf{k}, \tag{13}$$

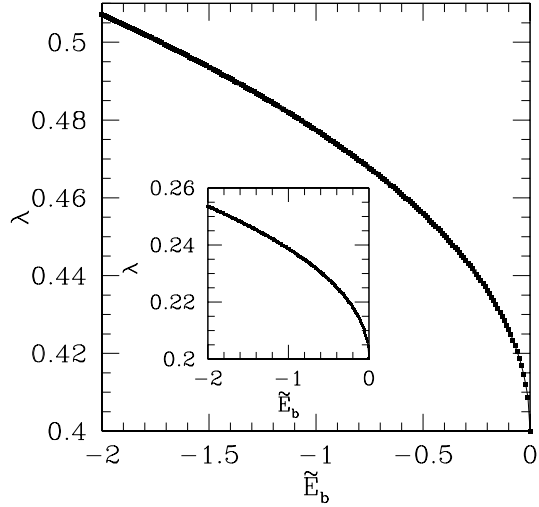

**FIGURE 1.** Interaction strength $\lambda$ (in units of $g_{2D}^{-1}$, where $g_{2D}$ is the two-dimensional density of states) as a function of $\tilde{E}_b = E_b(h_{\bar{z}}) + 2g_{\bar{z}\bar{z}}\mu_B h_{\bar{z}}$ (in units of $\varepsilon_F$) for $k_0 = 10k_F$ and $k_1 = k_F$ in the case of $\Gamma(\mathbf{k}) = h(k)\cos(\varphi)$. *Inset:* Same quantity in the case of $\Gamma(\mathbf{k}) = h(k)e^{i\varphi}$.

where $\tilde{E}_b = E_b(h_{\bar{z}}) + 2g_{\bar{z}\bar{z}}\mu_B h_{\bar{z}}$. After some algebra, and using the separability of the interaction potential, one finally obtains:

$$\frac{1}{\lambda} = \sum_{\mathbf{k}} \frac{|\Gamma(\mathbf{k})|^2}{2\varepsilon_{\mathbf{k}} - \tilde{E}_b}. \tag{14}$$

The dependence of $\lambda$ on $\tilde{E}_b$ is shown in Fig. 1 for the symmetry factors $\Gamma(\mathbf{k}) = h(k)\cos(\varphi)$ (leading to a $p_x$-symmetry order parameter) and $\Gamma(\mathbf{k}) = h(k)e^{i\varphi}$ (leading to a $p_x + ip_y$-symmetry order parameter). Observe that $\lambda$ remains positive for all values of $\tilde{E}_b < 0$, and does **not** change sign at the transition point $\tilde{E}_b^{(c)}$ (which corresponds to $\tilde{\mu} = 0$, as discussed below). This can be understood by noticing that $\lambda$ is simply the amplitude of the interaction potential in $k$-space (see Eq.(7)) and, therefore, is not the only factor responsible for the sign of $V_{\mathbf{kk}'}$. We would like to emphasize that the expression of $\lambda$ in terms of $\tilde{E}_b$ is just a convenient way of describing the theory in terms of the two-body energy $\tilde{E}_b$. We focus only on the case where $\tilde{E}_b$ is negative, i.e., when a two-body bound state appears, since this corresponds to the more physically interesting case, as shown below. By contrast, the corresponding situation in 3D $s$-wave systems is associated with a divergence and a change in sign of the scattering length $a_s$ when a two-body bound state appears. In this case, it is common (although strictly incorrect) to refer to the inverse scattering length $1/a_s$ as the *effective* two-body interaction, which would then change sign when $a_s \to \pm\infty$. It is better to associate this change with the appearance of a two-body bound state, when the potential is sufficiently attractive. (See, for example, the zero-range $s$-wave system studied in [7]).

Finally, the renormalized gap equation in terms of $\tilde{E}_b$ then takes the form

$$\sum_{\mathbf{k}} |\Gamma(\mathbf{k})|^2 \left[ \frac{1}{2\varepsilon_{\mathbf{k}} - \tilde{E}_b} - \frac{\tanh(\beta E_{\mathbf{k}}/2)}{2E_{\mathbf{k}}} \right] = 0. \tag{15}$$

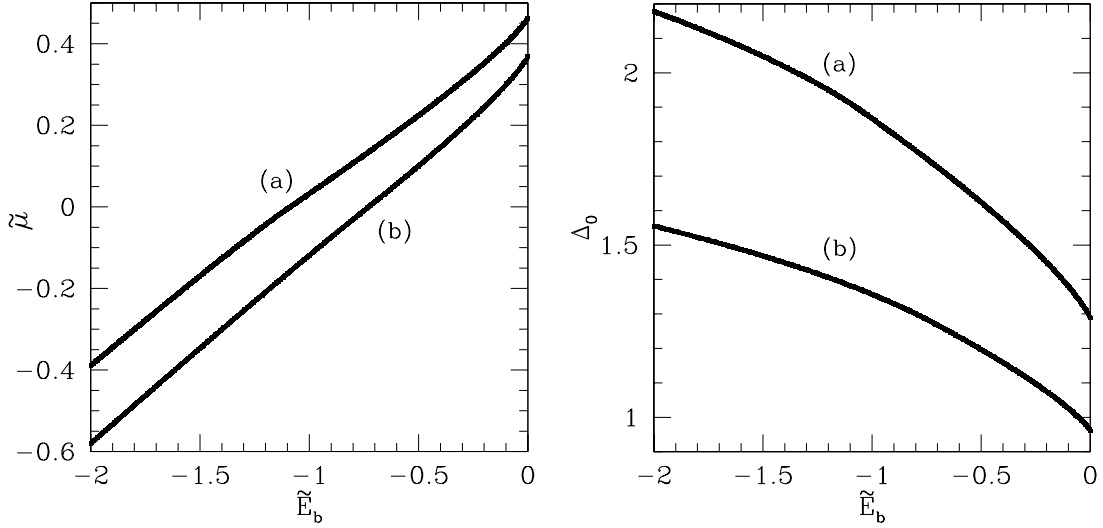

**FIGURE 2.** Universal plot (for any magnetic field $h_z$) of $\tilde{\mu} = \mu + g_{zz}\mu_B h_z$ and $\Delta_0$ as functions of $\tilde{E}_b = E_b(h_z) + 2g_{zz}\mu_B h_z$ (all quantities in units of $\varepsilon_F$) for $k_0 = 10k_F$ and $k_1 = k_F$ in the spin-polarized $p$-wave case with (a) $\Gamma(\mathbf{k}) = h(k)\cos(\varphi)$ and (b) $\Gamma(\mathbf{k}) = h(k)e^{i\varphi}$.

*Number Equation:* Using the relation $N = -\partial\Omega/\partial\mu$ and the saddle point approximation for the thermodynamic potential, $\Omega_0 = S_{\text{eff}}[\Delta_0]/\beta$, one can write the number equation as $N_0 = \sum_{\mathbf{k}} n_{\mathbf{k}}$, where the momentum distribution $n_{\mathbf{k}}$ is given by

$$n_{\mathbf{k}} = \frac{1}{2}\left[1 - \frac{\xi_{\mathbf{k}}}{E_{\mathbf{k}}}\tanh\left(\frac{\beta E_{\mathbf{k}}}{2}\right)\right]. \tag{16}$$

Thus, at $T = 0$, the saddle point and number equations reduce to $\sum_{\mathbf{k}}|\Gamma(\mathbf{k})|^2[(2\varepsilon_{\mathbf{k}} - \tilde{E}_b)^{-1} - (2E_{\mathbf{k}})^{-1}] = 0$ and $N_0 = \sum_{\mathbf{k}}(1 - \xi_{\mathbf{k}}/E_{\mathbf{k}})/2$, respectively. The solutions for $\Delta_0$ and $\tilde{\mu}$ at $T = 0$ as functions of the binding energy $\tilde{E}_b$ in the case of $p$-wave pairing symmetry are plotted in Fig. 2 for $k_0 = 10k_F$ and $k_1 = k_F$. The point $\tilde{\mu} = 0$ is achieved for the critical binding energy $\tilde{E}_b^{(c)} = -1.087\varepsilon_F$ (corresponding to $\Delta_0 = 1.906\varepsilon_F$) when the function $\Gamma(\mathbf{k}) = h(k)\cos(\varphi)$ is used in the interaction potential, and for $\tilde{E}_b^{(c)} = -0.729\varepsilon_F$ (corresponding to $\Delta_0 = 1.277\varepsilon_F$) in the case of $\Gamma(\mathbf{k}) = h(k)e^{i\varphi}$. Notice that very similar results are obtained with these two functions, for reasons that will be discussed later in the manuscript.

## 4. GAUSSIAN FLUCTUATIONS

We now investigate the effect of Gaussian fluctuations in the pairing field $\phi_{\mathbf{q}}(\tau)$ about the static saddle point value $\Delta_0$. Assuming $\phi_{\mathbf{q}}(\tau) = \Delta_0 \delta_{\mathbf{q},0} + \eta_{\mathbf{q}}(\tau)$ and performing an expansion in $S_{\text{eff}}$ to quadratic order in $\eta$, one obtains

$$S_{\text{Gauss}} = S_0[\Delta_0] + \frac{1}{2}\sum_q \underline{\eta}^\dagger(q)\mathbf{M}(q)\underline{\eta}(q), \tag{17}$$

where $S_0$ is the saddle point action, the vector $\underline{\eta}(q)$ is such that $\underline{\eta}^\dagger(q) = [\eta^*(q), \eta(-q)]$, and $q \equiv (\mathbf{q}, iq_m)$, where $iq_m = i2m\pi/\beta$ is a bosonic Matsubara frequency. The $2 \times 2$ matrix $\mathbf{M}(q)$ is the inverse fluctuation propagator.

The Gaussian fluctuation term in the effective action leads to a correction to the thermodynamic potential, which can be rewritten as $\Omega_{\text{Gauss}} = \Omega_0 + \Omega_{\text{fluct}}$, with $\Omega_{\text{fluct}} = \beta^{-1} \sum_q \ln \det[\mathbf{M}(q)]$. Therefore, using the relation $N = -\partial \Omega/\partial \mu$, one can write the corrected number equation as $N_{\text{Gauss}} = N_0 + N_{\text{fluct}}$, where $N_0$ is the saddle-point level number of particles given above, and

$$N_{\text{fluct}} = -\frac{\partial \Omega_{\text{fluct}}}{\partial \mu} = T \sum_{\mathbf{q}} \sum_{iq_n} \left[ \frac{-\partial (\det \mathbf{M})/\partial \mu}{\det \mathbf{M}(\mathbf{q}, iq_n)} \right]. \tag{18}$$

At low $T$, the Goldstone mode $\omega = c|\mathbf{q}|$ dominates the contribution to $N_{\text{fluct}}$, leading to

$$N_{\text{fluct}} \sim -\frac{L^2}{2\pi} \zeta(3) \frac{1}{c^3} \frac{\partial c}{\partial \mu} T^3, \tag{19}$$

which vanishes in the limit of $T \to 0$. Therefore, analogously to the three-dimensional $s$-wave case [8], Eq.(16) provides a very accurate description of the number equation near and at $T = 0$, thus confirming Leggett's suggestion [23]. However, it is well known that the same is not true near $T_c$, where the effects of temporal fluctuations are essential to describe the BEC regime [7]. The discussion of this interesting limit will be postponed to a future manuscript.

## 5. MOMENTUM DISTRIBUTION

The momentum distribution $n_{\mathbf{k}}$ given by Eq.(16), which at zero temperature reduces to $n_{\mathbf{k}} = (1 - \xi_{\mathbf{k}}/E_{\mathbf{k}})/2$, is plotted in Fig. 3 for $p$-wave pairing in the case where the symmetry factor $\Gamma(\mathbf{k}) = h(k)\cos(\varphi)$ is used. This corresponds to the situation where the quasiparticle excitation energy $E_{\mathbf{k}} = \left(\xi_{\mathbf{k}}^2 + \Delta_0^2 h^2(k)\cos^2(\varphi)\right)^{1/2}$ is gapless in the BCS regime and fully gapped in BEC regime. Notice that $n_{\mathbf{k}}$ becomes discontinuous when $\tilde{\mu}$ crosses zero, which coincides with the collapse of the two Dirac points to a single point $\mathbf{k} = 0$ and the appearance of a full gap to the addition of quasiparticles.

The use of the symmetry factor $\Gamma(\mathbf{k}) = h(k)e^{i\varphi}$ results in an angle-independent quasiparticle excitation energy $E_{\mathbf{k}} = \left(\xi_{\mathbf{k}}^2 + \Delta_0^2 h^2(k)\right)^{1/2}$, which is gapped on both the BCS and BEC sides of the transition line. Consequently, the momentum distribution will have no Dirac points, but will possess polar symmetry in $k$-space, as shown in Fig. 4. Notice, however, that the $\tilde{\mu} = 0$ momentum distribution develops a cusp at $\mathbf{k} = 0$ in this case, which is a consequence of the vanishing of $E_{\mathbf{k}}$ at $\mathbf{k} = 0$ when the chemical potential crosses the bottom of the band. This major rearrangement of $n_{\mathbf{k}}$ for both symmetry factors analyzed has a dramatic effect in the atomic compressibility $\kappa$, to be discussed next.

## 6. ATOMIC COMPRESSIBILITY

The isothermal atomic compressibility, defined as

$$\kappa = -\frac{L^2}{N^2} \left(\frac{\partial^2 \Omega}{\partial \mu^2}\right)_T = \frac{1}{n^2} \left(\frac{\partial n}{\partial \mu}\right)_T, \tag{20}$$

where $n = N/L^2$, develops a cusp when expressed in terms of $\tilde{E}_b$ (or $\mu$), and the first derivative of $\kappa$ with respect to $\tilde{E}_b$ is discontinuous at the critical point $\tilde{E}_b^{(c)}$ (that corresponds to $\tilde{\mu} = 0$), as shown in Fig. 5. This result is a consequence of the momentum distribution rearrangement when the chemical potential crosses the bottom of the band, as seen from the following alternative expression of $\kappa$ in terms of $n_{\mathbf{k}}$,

$$n^2 \kappa = \frac{2}{L^2} \sum_{\mathbf{k}} \frac{n_{\mathbf{k}}(1-n_{\mathbf{k}})}{E_{\mathbf{k}}} + \frac{\partial n_{\text{fluct}}}{\partial \mu}, \tag{21}$$

where $n_{\text{fluct}} = N_{\text{fluct}}/L^2$.

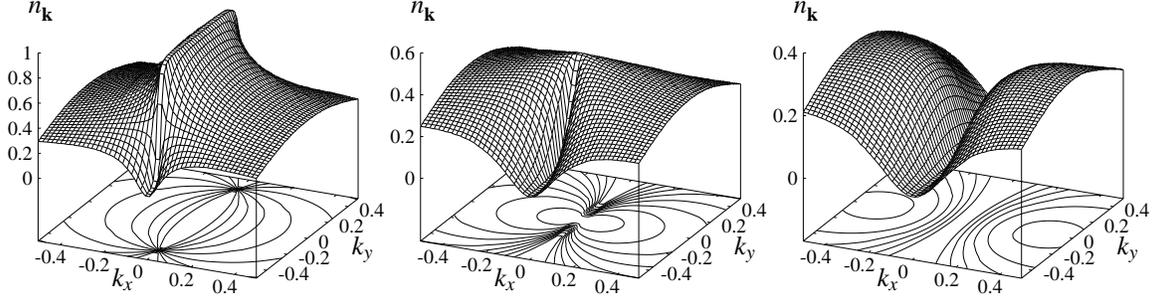

**FIGURE 3.** Plot of the momentum distribution $n_{\mathbf{k}}$ in the spin-polarized $p$-wave case with $\Gamma(\mathbf{k}) = h(k)\cos(\varphi)$ for (a) $\tilde{\mu} = 0.15\varepsilon_F$, (b) $\tilde{\mu} = 0$ and (c) $\tilde{\mu} = -0.15\varepsilon_F$. Notice the collapse of the two Dirac points when $\tilde{\mu}$ crosses zero. (Notice scale changes in figures.)

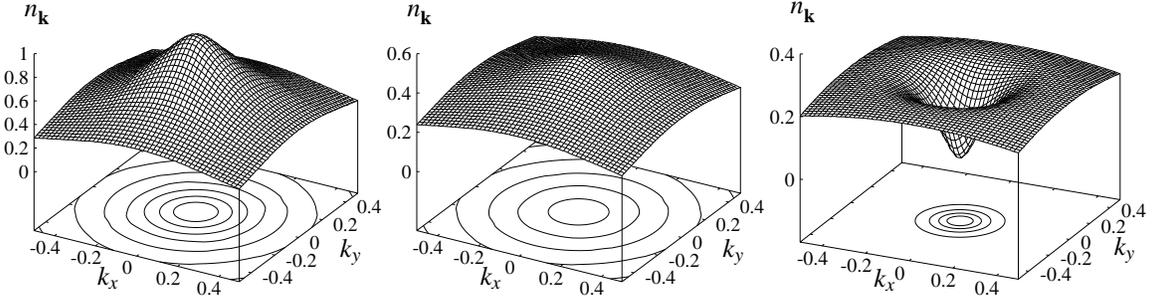

**FIGURE 4.** Plot of the momentum distribution $n_{\mathbf{k}}$ in the spin-polarized $p$-wave case with $\Gamma(\mathbf{k}) = h(k)e^{i\varphi}$ for (a) $\tilde{\mu} = 0.15\varepsilon_F$, (b) $\tilde{\mu} = 0$ and (c) $\tilde{\mu} = -0.15\varepsilon_F$. Notice the appearance of a cusp in $n_{\mathbf{k}}$ at $\mathbf{k} = 0$ when $\tilde{\mu}$ crosses zero. (Notice scale changes in figures.)

The non-analytic behavior of the atomic compressibility in the case of the symmetry function $\Gamma(\mathbf{k}) = h(k)\cos(\varphi)$ can be explained in terms of the collapse of the Dirac points toward $\mathbf{k} = 0$, together with the vanishing of $E_{\mathbf{k}}$ when $\tilde{\mu}$ crosses zero. On the other hand, the occurrence of a similar non-analyticity in $\kappa$ for the symmetry function $\Gamma(\mathbf{k}) = h(k)e^{i\varphi}$ can be related to the appearance of a cusp on the $\tilde{\mu} = 0$ momentum distribution at $\mathbf{k} = 0$, and again to the vanishing of $E_{\mathbf{k}}$ when $\tilde{\mu}$ crosses zero. This leads to the conclusion that the existence of Dirac points in the quasiparticle excitation spectrum is not necessary to produce a non-analytic behavior in the atomic compressibility, which can still occur even in the case of an angle-independent order parameter, as long as $h(k)$ vanishes as $\mathbf{k} \to 0$. In the $s$-wave case, however, $\kappa$ can be shown to be smooth for all values of $\tilde{E}_b$ [19]. This non-analytic behavior of the $p$-wave atomic compressibility, combined with the momentum distribution rearrangement, suggests the existence of a quantum critical point at $\tilde{\mu} = 0$.

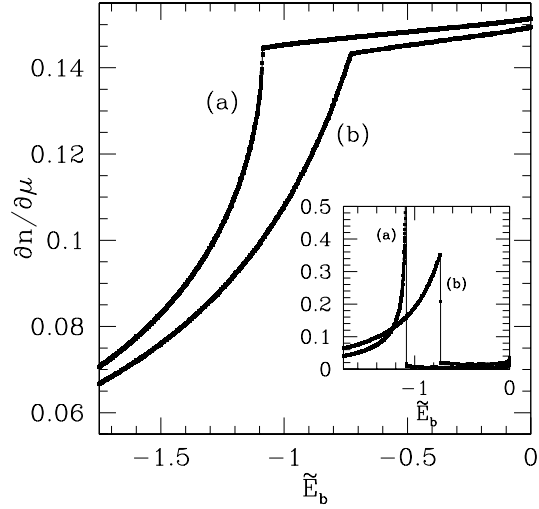

**FIGURE 5.** Plot of $\partial n/\partial \mu$ (in units of $k_F^2/4\pi\varepsilon_F$) and its first derivative with respect to $\tilde{E}_b$ (inset) as functions of $\tilde{E}_b$ in the case of spin-polarized p-wave pairing and $k_0 = 10k_F$ and $k_1 = k_F$, with (a) $\Gamma(\mathbf{k}) = h(k)\cos(\varphi)$ and (b) $\Gamma(\mathbf{k}) = h(k)e^{i\varphi}$.

## 7. SPIN SUSCEPTIBILITY

The phase transition discussed in the previous section also manifests itself in the spin susceptibility. The application of a small probe magnetic field $H_{\tilde{z}}$ along the same direction ($\tilde{z}$) of $\mathbf{h}$ generates the *spin* susceptibility response $\chi_{\tilde{z}\tilde{z}} = (-1/L^2)(\partial^2\Omega/\partial H_{\tilde{z}}^2)$, which can be rewritten in the case of spin-polarized atoms as

$$\chi_{\tilde{z}\tilde{z}} = -\frac{1}{L^2}g_{\tilde{z}\tilde{z}}^2\mu_B^2\frac{\partial^2\Omega}{\partial\mu^2} = g_{\tilde{z}\tilde{z}}^2\mu_B^2\frac{\partial n}{\partial\mu}. \tag{22}$$

Thus, the graph in Fig. 5 also represents a universal plot of $\chi_{\tilde{z}\tilde{z}}/g_{\tilde{z}\tilde{z}}^2\mu_B^2$ as a function of $\tilde{E}_b$.

## 8. SUPERFLUID DENSITY

We now turn our attention to the behavior of the low temperature superfluid density tensor $\rho_{ij}(T,\tilde{E}_b)$ as the critical value of the binding energy $\tilde{E}_b$ is crossed. This tensor is associated with phase twists of the superconductor order parameter [24] and is given by

$$\rho_{ij}(T) = \frac{1}{2L^2}\sum_{\mathbf{k}}\left[2n_{\mathbf{k}}\partial_i\partial_j\xi_{\mathbf{k}} - Y_{\mathbf{k}}\partial_i\xi_{\mathbf{k}}\partial_j\xi_{\mathbf{k}}\right], \tag{23}$$

where $n_{\mathbf{k}}$ is the momentum distribution, $Y_{\mathbf{k}} = (2T)^{-1}\text{sech}^2(E_{\mathbf{k}}/2T)$ is the Yoshida distribution, and $\partial_i$ denotes the partial derivative with respect to $k_i$. Notice that $\rho_{xx} = \rho_{yy} \equiv \rho$, while $\rho_{xy} = \rho_{yx} = 0$. In addition, notice that at $T = 0$, $\rho(0) = n/m$, such that $\partial\rho/\partial\mu = (1/m)\partial n/\partial\mu$ and $\partial\rho/\partial H_{\tilde{z}} = (1/m)\partial n/\partial H_{\tilde{z}}$.

Using our energy and momentum scales, we define the dimensionless quantity $\Delta\rho(T) \equiv m\rho(T)/n - 1$, which emerges naturally from the calculations. This quantity is shown in Fig. 6 as a function of temperature for different values of the binding energy, in the case of the symmetry function $\Gamma(\mathbf{k}) = h(k)\cos(\varphi)$. The linear behavior of $\Delta\rho(T)/T^2$ for values of $\tilde{E}_b$ that correspond to $\tilde{\mu} > 0$ indicates a $T^3$ dependence of the superfluid density on temperature on the BCS side of the transition. This behavior is in fact confirmed by our analytical calculation of $\Delta\rho(T)$ at low temperatures and in the case of short range interactions ($k_0 \to \infty$). In the BCS limit, we found $\Delta\rho(T) \sim CT^3$, with the coefficient $C$ weakly dependent on $\tilde{E}_b$. This power-law behavior reflects the nodal (gapless) structure of the $p$-wave excitation spectrum. In the BEC limit, we obtained $\Delta\rho(T) \sim \exp(-|\tilde{\mu}|/T)$, the exponential behavior reflecting the appearance of a full gap to the addition of quasiparticles for $\tilde{\mu} < 0$. Fig. 6 also shows (inset) the zero temperature slope of $\Delta\rho(T)/T^2$ as a function of the binding energy $\tilde{E}_b$, which is clearly discontinuous at the critical point $\tilde{E}_b^{(c)} = -1.087\varepsilon_F$.

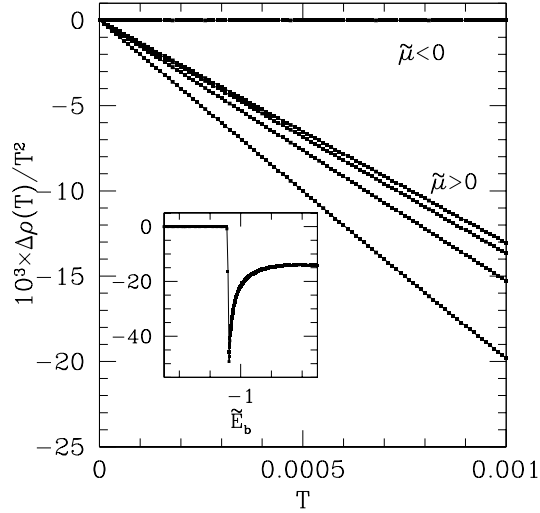

**FIGURE 6.** Plot of $\Delta\rho(T)/T^2$ (in units of $\varepsilon_F^{-2}$) as a function of temperature (in units of $\varepsilon_F$) for various values of the binding energy $\tilde{E}_b$, in the case of the symmetry factor $\Gamma(\mathbf{k}) = h(k)\cos(\varphi)$ with $k_0 = 10k_F$ and $k_1 = k_F$. *Inset:* Zero-temperature slope of $\Delta\rho(T)/T^2$ (in units of $\varepsilon_F^{-3}$) as a funcion of $\tilde{E}_b$.

These results further confirm the existence of a quantum phase transition along the BCS-to-BEC evolution as a function of interaction strength (binding energy) in the case of $p$-wave spin-polarized atoms. It is important to point out that there is an additional contribution to $\rho(T)$ due to Goldstone modes (underdamped for $\tilde{\mu} > 0$ due to Landau damping, but not damped for $\tilde{\mu} < 0$ due to a full gap in the quasiparticle spectrum). In our formulation, this contribution comes as a next order correction, and has the form $\rho_G(T) = -AT^3/c^4$ at low temperatures, where $A$ is a function of $\tilde{E}_b$ and $c$ is the speed of sound. However, for $|\tilde{\mu}| \leq 0.4\varepsilon_F$ (which corresponds to $-2\varepsilon_F \leq \tilde{E}_b \leq 0$), $A$ is essentially unchanged, and $\rho_G(T)$ does not contribute to the discontinuity in the slope of $\Delta\rho(T)/T^2$ shown in Fig. 6.

In the case of the symmetry factor $\Gamma(\mathbf{k}) = h(k)e^{i\varphi}$, it is clear that the superfluid density will have an exponential behavior on both sides of the transition line, vanishing as $\rho(T) \sim B_1 \exp(-\Delta_{\mathbf{k}_F}/T)$ on the BCS side, and as $\rho(T) \sim B_2 \exp(-|\tilde{\mu}|/T)$ on the BEC side. This occurs because the quasiparticle excitation spectrum is gapped in both limits. Therefore, the quantum phase transition that we described will not manifest itself explicitly in this case through the temperature dependence of the superfluid density. However, this fact does not preclude the pre-factors $B_1$ and $B_2$, which are functions of $\tilde{\mu}$ only, from having different behaviors for $\tilde{\mu} > 0$ and $\tilde{\mu} < 0$. In fact, this quantum phase transition will still be explicit in the $T = 0$ behavior of $\rho(T = 0) = n/m$, since $\partial \rho / \partial \mu = (1/m) \partial n / \partial \mu$ is directly proportional to the atomic compressibility $\kappa$.

## 9. KOSTERLITZ-THOULESS TRANSITION

Before summarizing our results, it is worth considering the possibility of a Berezinskii-Kosterlitz-Thouless (BKT) phase transition [25, 26] in the system under consideration. We start with the observation that the existence of a finite atomic transfer energy $t_z$ causes the system to be non-two-dimensional. One is then left with the question of whether such a finite $t_z$ can lead the system towards an effectively three-dimensional behavior (in which case it should be better described as quasi-two-dimensional), or if $t_z$ is really small enough to guarantee an essentially two-dimensional behavior, such that a BKT transition would be possible. In order to answer this question, we first derived an approximate expression for the atomic transfer energy $t_z$ (based on a WKB calculation), and obtained

$$t_z \approx \frac{1}{2} \hbar \omega_0 \exp\left[-\pi \sqrt{\frac{1}{E_r}\left(V_0 - \frac{\hbar \omega_0}{2}\right)}\right], \tag{24}$$

where $\hbar \omega_0 = \sqrt{2V_0 E_r}$, and $E_r = \hbar^2 k_z^2 / 2m$ is the recoil energy, where $\lambda_z = 2\pi/k_z$ is the wavelength of the laser. In order to estimate a numerical value for $t_z$, we used typical experimental parameters for the laser wavelength ($\lambda_z = 852$ nm), particle density ($n_{2D} = 2.50 \times 10^7 \text{cm}^{-2}$ after conversion to the quasi-two-dimensional situation), and fermion mass ($m = 6.49 \times 10^{-26}$Kg for $^{40}$K). It is clear that the parameter $t_z$ can be tunned depending on the laser intensity (related to $V_0$). Typical maximum laser intensities in optical lattices [27] correspond to $V_0 = 20 E_r$; however, for our purposes we take a smaller intensity corresponding to $V_0 = 5 E_r$. With the help of these numbers, we obtained $t_z \approx 0.0822 \varepsilon_F$. Based on this result, and on the fact that near $\tilde{\mu} = 0$ the system is already close to the Bose limit (from the critical temperature point of view), we calculated the Bose-Einstein condensation temperature $T_{BE}$ (using $k_B = 1$) for our highly-anisotropic three-dimensional system (see the Appendix for details),

$$T_{BE} \approx 0.487 \varepsilon_F \left(\frac{t_z^{(B)}}{\varepsilon_F}\right)^{1/3}, \tag{25}$$

where $t_z^{(B)} = t_z^2 / E_b$ is the effective transfer energy of the composite boson (bound state) along the lattice direction. In the case of $p_x$-symmetry, one finds $T_{BE} = 0.0896 \varepsilon_F$, while for $p_x + ip_y$-symmetry, $T_{BE} = 0.102 \varepsilon_F$ at the point of interest. On the other hand, the BKT temperature $T_{BKT}$ [25, 26] of a strictly two-dimensional system ($t_z = 0$) can be related to the superfluid density $\rho(T)$ via the self-consistent equation [28]

$$\frac{\pi}{2} \hbar^2 \rho(T_{BKT}) = T_{BKT}. \tag{26}$$

Solving this equation numerically, we obtained an upper bound for the BKT temperature of $T_{BKT}^{(\max)} = 0.0517\varepsilon_F$ for $p_x$-symmetry, and $T_{BKT}^{(\max)} = 0.0809\varepsilon_F$ for $p_x + ip_y$-symmetry. In both cases, the conclusion that $T_{BE} > T_{BKT}^{(\max)}$ implies that the tunneling rate is indeed large enough to drive the system toward an effectively three-dimensional behavior. In the language of renormalization group, this is equivalent to saying that $t_z$ is sufficient to cause the system to converge to a three-dimensional critical point rather than to a two-dimensional critical point, such that the BKT transition does not occur. It is important to note, however, that one could in principle lead the system toward the BKT regime by using different experimental parameters. In particular, by increasing the intensity of the laser, one could cause $t_z$ to be further reduced, so that one could have $T_{BE} < T_{BKT}$, and the BKT transition would become possible. Although this limit certainly deserves further investigation, we showed that one can always keep the system away from it by using appropriate experimental parameters. For instance, by using a laser beam whose intensity is low enough, one can cause the tunneling rate between lattice sites to become non-negligible, so that the system is essentially quasi-two-dimensional (still satisfying $t_z \ll \varepsilon_F$), and the BKT limit is not attainable.

## 10. SUMMARY

We proposed the existence of a quantum phase transition in the BCS-to-BEC evolution of $p$-wave fully spin-polarized Fermi gases as a function of the two-body bound state energy. Two different types of order parameter symmetries were analyzed: $p_x$ and $p_x \pm ip_y$. In both cases, we have shown that the momentum distribution undergoes a major rearrangement in $k$-space at a critical value of the binding energy, which leads to a non-analytic behavior of the atomic compressibility and spin susceptibility of the gas. Furthermore, in the case of $p_x$-symmetry, the low temperature superfluid density of the system was shown to change dramatically as the critical point is crossed, with a zero-temperature slope that is discontinuous at a critical binding energy $\tilde{E}_b^{(c)}$.

We conclude by suggesting that this phase transition may be observable in traps of $^6$Li and $^{40}$K gases which exhibit $p$-wave Feshbach resonances [17, 18]. The occurrence of this phase transition may be investigated through the direct measurement of the atomic compressibility, spin susceptibility or superfluid density as functions of binding energy or magnetic field.

## ACKNOWLEDGMENTS

We would like to thank NSF (Grant No. DMR 0304380) for financial support, and Chandra Raman and Tony Leggett for references and discussions.

## APPENDIX

The expression for the Bose-Einstein condensation temperature $T_{BE}$ given in Eq.(25) is calculated using a composite boson dispersion

$$\varepsilon_B(\mathbf{q}) = E_0^{(B)} + \frac{\hbar^2}{2m_B}(q_x^2 + q_y^2) - 2t_z^{(B)}\cos(q_z a_z), \tag{27}$$

where $t_z^{(B)} = t_z^2/E_b$ is the boson transfer energy between neighboring lattice sites, and $a_z = \lambda_z/2$ is the optical lattice spacing. From the expansion of $\varepsilon_B(\mathbf{q})$ for small $q_z a_z$, one obtains the effective mass $m_z^{(B)} = \hbar^2/(2t_z^{(B)} a_z^2)$ along the optical lattice direction ($z$). Using the resulting quadratic dispersion, the Bose-Einstein condensation temperature $T_{BE}$ for the anisotropic three-dimensional system can be expressed in terms of the boson mass $m_B = 2m_F$, the boson density $n_B = n_F^{(3D)}/2$ (where $n_F^{(3D)}$ is the fermion density in 3D), and the mass anisotropy ratio $\alpha = (m_z^{(B)}/m_B)^{1/2}$, as

$$T_{BE} = T_{BE}^{(\text{iso})} \alpha^{-2/3}. \tag{28}$$

In this expression,

$$T_{BE}^{(\text{iso})} = \frac{2\pi\hbar^2}{m_B}\left[\frac{n_B}{\zeta(3/2)}\right]^{2/3} = 0.137\varepsilon_F^{(3D)} \tag{29}$$

is the Bose-Einstein condensation temperature for the 3D isotropic case (where $\alpha = 1$) in the spin polarized (single pseudo-spin state) case. Here, $\varepsilon_F^{(3D)} = \hbar^2 k_{F,3D}^2/2m_F$ is the Fermi energy of the isotropic 3D Fermi gas. Notice that this result is different from the case of two-spin (or two-pseudo-spin) states, where $T_{BE}^{(\text{iso})} = 0.218\varepsilon_F^{(3D)}$ because of the larger spin (pseudo-spin) degeneracy. In addition, notice that $T_{BE}$ is always smaller than $T_{BE}^{(\text{iso})}$, since our result is valid only for $\alpha \geq 1$.

As can be seen in Eqs. (28) and (29), $T_{BE}$ is expressed in terms of $\varepsilon_F^{(3D)}$. However, it is to our advantage to express $T_{BE}$ in terms of the 2D Fermi energy $\varepsilon_F$ (used throughout the manuscript), because we would like to compare $T_{BE}$ with $T_{BKT}$, which is naturally expressed in units of $\varepsilon_F$. In order to do this, it is necessary to relate $\varepsilon_F^{(3D)}$ and $\varepsilon_F$. In the spin polarized (single pseudo-spin) case, this is achieved by noting that the 3D density $n_F^{(3D)} = (k_{F,3D}^2)^{3/2}/6\pi^2$ is related to the 2D density $n_F = k_F^2/4\pi$ via the relation $n_F^{(3D)} = n_F/a_z$. This procedure leads to the relation $\varepsilon_F^{(3D)} = \varepsilon_F(3\pi/2k_F a_z)^{2/3}$. Using this relation and the expression for $\alpha = (\varepsilon_F/2t_z^{(B)})^{1/2}/k_F a_z$ in terms of the 2D Fermi energy into Eq. (28), one directly obtains Eq. (25). Our results for the 3D anisotropic $T_{BE}$ are in agreement with previous estimates. [29]